\pdfoutput=1
\documentclass[twocolumn,epjc3]{svjour3}
\usepackage{amsmath,graphicx}
\usepackage{multirow}
\usepackage{booktabs}
\usepackage{color}
\usepackage{soul}
\usepackage{txfonts}
\usepackage{cuted}

\newcommand{\prl}{Phys. Rev. Lett.}
\newcommand{\prd}{Phys. Rev. D}
\newcommand{\jcap}{JCAP}
\newcommand{\aap}{Astron. \& Astrop.}

\def\apj{The Astrophysical Journal}
\def\mnras{Monthly Notices of the Royal Astronomical Society}

\definecolor{Green}{RGB}{255,0,0}

\RequirePackage[T1]{fontenc}

\smartqed  

\RequirePackage{graphicx}
\RequirePackage{mathptmx}      
\RequirePackage{flushend}
\RequirePackage[numbers,sort&compress]{natbib}
\RequirePackage[colorlinks,citecolor=blue,urlcolor=blue,linkcolor=blue]{hyperref}

\journalname{Eur. Phys. J. C}

\begin{document}

\title{
Dynamical and observational properties of weakly Proca-charged black holes
}
\author{Abylaikhan Tlemissov\thanksref{e1,addr1}
        \and Arman Tursunov\thanksref{e2,addr1,addr2}\and
        Ji\v{r}\'{i} Kov\'{a}\v{r}\thanksref{e3,addr1} \and Zden\v{e}k Stuchlík\thanksref{e4,addr1}
}
\thankstext{e1}{e-mail: tle0002@slu.cz}
\thankstext{e2}{e-mail: arman.tursunov@physics.slu.cz}
\thankstext{e3}{e-mail: jiri.kovar@fpf.slu.cz}
\thankstext{e4}{e-mail: zdenek.stuchlik@physics.slu.cz}

\institute{Research Centre for Theoretical Physics and Astrophysics, Institute of Physics, Silesian University in Opava,\\ Bezru\v{c}ovo n\'{a}m. 13, CZ-746\,01, Opava, Czech Republic\label{addr1} \and Max Planck Institute for Radio Astronomy, Auf dem H{\"u}gel 69, Bonn D-53121, Germany\label{addr2}
}

\date{Received: date / Accepted: date}

\maketitle

\begin{abstract}
The simplest approach to include a mass into the electromagnetic vector potential is to modify the Einstein-Maxwell action to the Einstein-Proca form. 
There are currently no exact analytical solutions for this scenario. However, by using perturbation theory, where both the Proca mass and the black hole charge are small parameters, it is possible to find an exact analytical solution. In this solution, the metric tensor remains unchanged, but the vector potential deviates from the Coulomb potential. In particular, even if the 
Proca mass is limited by the value $m_{\gamma}<10^{-48}\text{g}$, which is 
the current experimental upper limit for photon mass, it makes a significant contribution to the dynamical equations. 
In this paper, we study the motion of neutral and charged particles in the vicinity of a weakly Proca-charged black hole, and test the observational implications of the solution of the Einstein-Proca equations for gravitational bending, the black hole shadow, and the fit to the orbits of the Galactic center flares observed by the near-infrared GRAVITY instrument. 
We find that only extremely cold photons, which are likely scattered before reaching a distant observer, could reveal the non-zero photon mass effect through the black hole shadow. For the Galactic center flare analysis we obtained constraints on the dimensionless Proca parameter to $\mu \leq 0.125$ for the electric interaction parameter in the range $-1.1 < {\cal Q} < 0.5$, which can be potentially tested by future GRAVITY flare astrometry. Since the Proca parameter is coupled to the black hole mass, the effect of the Proca charge becomes more pronounced for supermassive black holes compared to stellar-mass objects. Our perturbative treatment remains valid essentially up to the horizon, with divergences appearing only in the immediate near-horizon region, where a fully non-perturbative analysis would be required. 
\end{abstract}

\section{INTRODUCTION}
The equation of motion for a free massive vector field was first proposed by Proca \cite{Proca:1936fbw}. This model is currently considered to be the simplest way to include the mass of the vector field, including the electromagnetic field. Therefore, the most famous and straightforward approach to incorporate the mass of a photon $m_\gamma$ in general relativity is to modify the Einstein-Maxwell Lagrangian density into Einstein-Proca with broken local gauge $U(1)$ symmetry. On the other hand, the no-hair theorem of black hole physics states that black holes can be described by only three parameters, the mass, spin, and charge. Indeed, the no-hair theorem poses a theoretical challenge to defining the mass of a photon. As argued by Bekenstein \cite{PhysRevD.5.1239}, in the case of a Schwarzschild black hole,  static massive vector fields cannot be used as an external source of black holes. Nevertheless, a comprehensive mathematical proof for a universal no-hair theorem remains elusive, and within mathematical circles, it is often referred to as the ``no-hair conjecture". 
In this paper we show that if the massive vector field is weak, one can incorporate it into the equations without violating the no-hair theorem.

There have been several experimental efforts to search for deviations from the $U(1)$-symmetry, in attempts to set an upper limit on the photon mass. 
The tests, mainly proposed in the 1960s, were based on e.g. probing of Coulomb's inverse square law at small distances, resulting in $m_{\gamma}\leq 1.6\times10^{-47}\text{g}$ \cite{1971PhRvL..26..721W}, analyzing astrophysical magnetic field data at large scales, giving $m_{\gamma}\leq 8\times10^{-49}\text{g}$ \cite{1971RvMP...43..277G,1975PhRvL..35.1402D,1994PhRvL..73..514F}), and studying the mechanical stability of magnetized
gases, leading to $m_{\gamma}\leq 3\times10^{-60}\text{g}$ 
\cite{1976SvPhU..19..624C}. However, the approach proposed by  \cite{1976SvPhU..19..624C} critically depends on many assumptions, and the reliability of this result remains somewhat unclear.  One of the first estimations, in which the photon mass has been limited was due to the effect of space curvature \cite{1973PhRvD...8.2349L},  exploiting the gravitational deflection of electromagnetic radiation. However, this method gives rather looser constraints compared to the other methods, namely $m_{\gamma}\leq 10^{-40}\text{g}$. 

More recently, robust limits on the photon mass have been obtained by 
using the observed dispersion measures from two samples of extragalactic pulsars \cite{Wei:2018pyh}. The limits of $m_{\gamma}\leq 1.51\times10^{-45}\text{g}$ ($m_{\gamma}\leq2.42\times10^{-45}\text{g}$) have been estimated at the confidence level of 68\% (95\%) for the sample of 22 radio pulsars in the Large Magellanic Cloud and $m_{\gamma} \leq
1.58 \times 10^{-45} \text{g}$ ($m_{\gamma} \leq 2.43 \times 10^{-45} \text{g}$) for the other sample of 5 radio pulsars in the Small Magellanic Cloud. 

A possible connection between the electrodynamics of massive photons and the dark energy has been explored by \cite{Kouwn:2015cdw}, obtaining a nonvanishing photon mass of the order of $\sim 10^{-34}\text{eV}/c^2$ capable of mimicking the current cosmological observations and the $\Lambda \text{CDM}$ model, which is also a limit obtained from the uncertainty principle for the current age of the universe. 
Inclusion of the massive spin-1 particles can also be interesting in connection with the dark matter problem \cite{Holdom:1985ag,Arkani-Hamed:2008hhe,Pospelov:2008jd,Goodsell:2009xc}, in which the examination of self-gravitating structures originating from these particles and their dynamic behaviour plays a significant role \cite{Arias:2012az}. 
In \cite{Brito:2015pxa} it was shown that complex massive spin-1 particles, similar to massive spin-0 particles, have the ability to merge under their gravitational self-attraction into smooth, asymptotically flat energy concentrations.

Another interesting application of the Proca formalism can be in the description of the effective mass of photons gained along their propagation in a plasma \cite{Robles:2012zz}. Under certain physical circumstances, the dynamical equations for photons in the presence of plasma are equivalent to those for massive particles propagating in a vacuum with a velocity lower than $c$. The effective mass in that case depends on the parameters of an environment and is proportional to the plasma frequency. For realistic astrophysical plasma, the effective mass of a photon can be as large as $10^{-10}$ of the rest mass of an electron.

We organize the paper in the following way. In Section 2, we present the Einstein-Proca equation in dimensionless form and solve it in the weak charge approximation. We also discuss the no-hair theorem and provide arguments on why it is not applicable in this approximation. In Section 3, we investigate the motion of charged particles around a black hole in the presence of a weak Proca field.  %
We focus on the study of the existence of circular orbits, in general. In Section 4, we discuss observational properties of weakly Proca-charged black holes focusing on two independent gravity tests: the gravitational lensing and fitting of the data from the observations of the Galactic center supermassive black hole flares. We analyse the deflection of a massive particle (photon) directly near a black hole and present a general expression for the size of the shadow of a black hole depending on the energy/velocity of a photon with non-zero mass. Our main results are discussed in Section 5. 

\section{Einstein-Proca theory and perturbation solution}\label{sec_2}
Classical Proca field minimally coupled with gravity is described by the action
\begin{equation}
	\mathcal{S}=\int {\rm d}^4x\sqrt{-g}\left(\frac{R}{2k}-\frac{1}{4\mu_0} F_{\mu\nu}F^{\mu\nu}-\frac{\mu^2}{2\mu_0}A_{\mu}A^{\mu}\right), 
	\label{Proca_Lagrangian}
\end{equation}
where $R$ is a scalar curvature,  $A_{\mu}$ is a vector potential, and $F_{\mu\nu}=\partial_{\mu}A_{\nu}-\partial_{\nu}A_{\mu}$ is a field tensor with the constants $\mu_0=1/c^2\epsilon_0$ and $k=8\pi G/c^4$. The Proca parameter $\mu=m_{B}c/\hbar$ is an inverse Compton wavelength of a boson of mass $m_B$. The Proca action holds the usual Lorentz invariance, but abandons the $U(1)$ local gauge invariance due to the
additional mass terms. The Proca field satisfies the equation 
 \begin{equation}
     \nabla_{\mu}F^{\mu\nu}=\mu ^2A^{\nu}.
     \label{Proca_eq_motion}
 \end{equation}
After varying  action \eqref{Proca_Lagrangian}, the field equation reads 
\begin{equation}   
R_{\mu\nu}=k\left(T_{\mu\nu}-\frac{1}{2}g_{\mu\nu}T\right), 
    \label{Einstein_eq}
\end{equation}
with the energy-momentum tensor in the form 
\begin{equation}
\begin{split}
\mu_0T_{\mu\nu}=\mu_0T_{\mu\nu}=&F_{\mu\sigma}F_{\nu}^{\hspace{1.5mm}\sigma}+\mu^2A_{\mu}A_{\nu}\\
&-g_{\mu\nu}\left(\frac{1}{4}F_{\sigma\delta}F^{\sigma\delta}-\frac{1}{2}\mu^2A_{\sigma}A^{\sigma}\right). 
\end{split}    
\end{equation}
In this paper we consider spherically symmetric case with the metric 
\begin{equation}
 {\rm d}s^2=-f(r)c^2{\rm d}t^2+\frac{1}{h(r)}{\rm d}r^2+r^2({\rm d}\theta^2+\sin^2{\theta}{\rm d}\phi^2).
 \label{Proca_metric}
\end{equation}
Equations for the functions $h(r)$, $f(r)$, and $A(r)$ can be derived by using the chosen static ansatz $F_{rt}=-A'(r)$ and written in the form
\begin{eqnarray}
h&=&\frac{k r^2\mu^2\epsilon_0A^2+2f}{2f+2rf'+kr^2\epsilon_0A'^2},
\label{h_equation}\\
\frac{2A'}{r}+A''&=&\mu^2 \frac{A}{h}\left(1+\frac{k\epsilon_0rAA'}{2f}\right),\label{A_equation}\\
    \frac{2f'}{r}+f''&=&k\epsilon_0A'^2+k\epsilon_0\mu^2\frac{A^2}{h}\left(2+\frac{rf'}{2f}\right).
    \label{f_equation}  
\end{eqnarray}
Here primes represent the derivatives with respect to the radial coordinate $r$. The vector potential {\bf $A_{\mu}$} then satisfies the relation
\begin{equation}
    h^2\left(\frac{f}{h}\right)'=k\epsilon_0r\mu^2 A^2. 
    \label{f/h_eq}
\end{equation}
There is no exact solution of equations \eqref{h_equation}, \eqref{A_equation} and \eqref{f_equation}. However, assuming naked singularity spacetimes, numerical solutions were considered in paper \cite{Obukhov:1999ed}. The asymptotic behavior of functions,  $f\rightarrow 1$ and $h\rightarrow 1$, yields $A=0$ for $r\rightarrow\infty$. Another flat spacetime behaviour appears if we put $f=h=1$ into \eqref{A_equation}, giving the approximation $A\approx e^{-\mu r}/r$, which rapidly goes to zero for $r>1/\mu$.

Further we briefly discuss the no-hair theorem based on works \cite{2017JCAP...08..024H}, and show that it does not apply to the weakly Proca-charged black holes. We assume that at the event horizon, $r=r_{h}$, the metric functions vanish, giving $f(r_{h})=h(r_{h})=0$. This implies that the functions $f$ and $h$ can be expressed as $f=\sum_{i=1}f_{i}(r-r_{h})^{i}$ and $h=\sum_{i=1}h_{i}(r-r_{h})^{i}$ respectively, in the vicinity of the horizon. Since, in this scenario, the left-hand side of the equation takes values of $\left(f_{2}h_{1}-h_{2}f_{1}\right)\left(r-r_{h}\right)^2+\mathcal{O}\left(\left(r-r_{h}\right)^3\right)$, we can conclude that the vector potential near the horizon must be proportional to
\begin{equation} A\approx\sqrt{\frac{f_{2}h_{1}-h_{2}f_{1}}{k \epsilon_{0}r_{h}\mu^2}}\left(r-r_{h}\right).
\end{equation}
According to Bekenstein \cite{PhysRevD.5.1239}, if the potential vanishes at the horizon and at infinity, then it should vanish everywhere outside the horizon. %
After some mathematical manipulations with the Proca equation \eqref{A_equation}, we get 
\begin{equation}
    \left[\frac{r^2AA'}{\sqrt{g}}\right]'=\frac{r^2}{\sqrt{g}}\left(A'^2+\frac{\mu^2A^2}{h}\right), 
\end{equation}
where $g=f/h$. Next, it is convenient to use the substitution $A=a/r$, which gives
\begin{equation}
\label{de}
     \left[\frac{a}{\sqrt{g}}\left(a'-\frac{a}{r}\right)\right]'=\frac{1}{\sqrt{g}}\left(\left(a'-\frac{a}{r}\right)^2+\frac{\mu^2a^2}{h}\right).
\end{equation}
After integrating both sides of the above equation from the horizon to spatial infinity, one can see that the left-hand side of equation (\ref{de}) equals to zero. The integrand on the right-hand side is positively defined, since $h$ is positive outside the horizon. Therefore, the only regular solution appears when $a=0$ or, equivalently, $A=0$.

Let us now define the dimensionless parameters $x=\mu r$ and $u=A/s$ with 
\begin{equation}
s=\frac{Q\mu}{\epsilon_0},\hspace{1cm} \epsilon=\frac{kQ^2\mu^2}{\epsilon_0}. 
\label{small_parameter}
\end{equation}
Here, $\epsilon$ represents a small perturbation parameter that is dimensionless, while $s$ is a constant that arises from electromagnetic theory. Using a weak charge approximation, $\epsilon \sim 0$, it can be shown that equation $\eqref{f/h_eq}$ does not contribute to the equations of motion and can be ignored. This is because, in the limit of a weak charge, the solutions for the metric are in the form of a Schwarzschild metric where $f=h$. Thus, the left-hand side of this equation equals to zero, and the right-hand side of the equation also becomes zero due to small charge.

The system of equations \eqref{A_equation}-\eqref{f_equation} in weakly charged approximation, $\epsilon\sim 0$, reduces to the form
\begin{eqnarray}
f\left(\frac{2f'}{x}+f''\right)&=&0,
\\
f\left(\frac{2u'}{x}+u''\right)&=&u\left(f+xf'\right)\label{u_eq}
\end{eqnarray}
where the right-hand side of equation \eqref{u_eq} appears purely due to Proca theory. Solution of these equations can be written in the form
\begin{eqnarray}
    f&=&h=1-\frac{C}{\mu r},
    \label{f_h_function_sol}
    \\
    A&=&s u=\frac{Qe^{C-\mu r}}{4\pi \epsilon_{0}r}U\left[\frac{C}{2},0,2(\mu r-C)\right].
    \label{u_A_t_function_sol}
\end{eqnarray}
Here $f$ and $h$ stand for the so-called lapse functions, $U(a,b,c)$ is the Tricomi Hypergeometric function, and we get back to radial coordinate, $r$, instead of the dimensionless parameter, $x$. Although the Schwarzschild term was neglected in the zero term in the paper \cite{2014arXiv1406.0497V}, i.e. $r\rightarrow \infty$,  we can retain the term proportional to $C/x$ here. Comparing the lapse functions, $f$ and $h$, with the Schwarzschild lapse function, we find 
\begin{equation*}
    C=\frac{2GM\mu}{c^2}.
\end{equation*}
Note that for $Q=1$ the solutions are exactly the same as in the Einstein-Cartan-Proca theories \cite{1984NCimB..80...62G,1995GReGr..27.1259G}, in which massive photons are interpreted as torsion quanta. From the properties of the function $U(0,0,z)=1$ we can see that in the absence of gravity the solutions for the vector potential changes to the Yukawa potential, and in the absence of the parameter $\mu$, we recover the Coulomb law.

In the geometrized unit system, $G=c=\frac{1}{4\pi\epsilon_0}=1$, in which we define the following dimensionless quantities 
\begin{equation*}
\begin{split}
    &\tilde{\mu}=\mu M,\quad \tilde{Q}=\frac{Q}{M},\quad \tilde{r}= \frac{r}{M},\\
    &\tilde{L}=\frac{L}{M},\quad \tilde{\tau}= \frac{\tau}{M},\quad \tilde{t}=\frac{t}{M}, 
\end{split}
\end{equation*}
the relations for $f$, $h$ and $A_t$ in the case of weakly Proca-charged black hole can be written in the form 
\begin{eqnarray}
       h&=&f,\\
    f&=&1-\frac{2}{\tilde{r}},\\
    A_{t}&=&-\frac{\tilde{Q}e^{-\tilde{\mu}\left(\tilde{r}-2\right)}}{\tilde{r}}U\left[\tilde{\mu},0,2\tilde{\mu}(\tilde{r}-2)\right],
    \label{vec_pot_eq}
\end{eqnarray}
exhibiting an event horizon located at $\tilde{r}=2$. However, the zero approximation of the vector potential does not turn to zero everywhere, as expected. From the property of hypergeometric functions, however, we find that a solution for the vector potential does exist outside the horizon $\tilde{r}\geqslant2$, and interior solution $\tilde{r}<2$ does exist only for $\tilde{\mu}=0$, which is equivalent to a weakly charged approximation of the Reissner–Nordstr\"{o}m black 
hole \cite{Tursunov:2021jjf}. 

At the horizon, the vector potential takes the value
\begin{equation*}
    A_{t}(\tilde{r}=2)=-\frac{\tilde{Q}}{2\Gamma(\tilde{\mu}+1)}, 
\end{equation*}
and differs from the Coulomb value by the factorial of $1/\tilde{\mu}$, and for large values of the mass parameter, $\tilde{\mu}$, it screens the electric field. At large distances, the potential reduces to the form
\begin{equation}
A_{t}(\tilde{r}\rightarrow\infty)=-\left(\frac{e^2}{2\tilde{\mu}}\right)^{\tilde{\mu}}\frac{\tilde{Q}e^{-\tilde{\mu} \tilde{r}}}{\tilde{r}^{1+\tilde{\mu}}}, 
\label{assymptotitc_potential}
\end{equation}
where we used the asymptotic behaviour $\lim_{z\rightarrow\infty} U(a,b,z)=z^{-a}$, which decreases faster than the Yukawa potential due to gravitational couplings.

To find out the contribution of $\epsilon$ to the exact solution, it is convenient to expand the functions $f$ and $u$ in a series by degrees of $\epsilon$ as follows:
\begin{equation}
    u=u_{0}+\epsilon u_{1}+\epsilon^2 u_{2}+\dots, \quad f=f_{0}+\epsilon f_{1}+\epsilon^2 f_{2}+\dots.
    \label{u_f_expandsion}
\end{equation}

In this scenario, the solutions \eqref{f_h_function_sol} and \eqref{u_A_t_function_sol} serve as the zero-orders terms in the series expansion of \eqref{u_f_expandsion}. 
The appendix provides details on how higher-order terms can be calculated. However, it has been demonstrated that, for the case of a weakly Proca-charged black hole, these higher-order terms can be safely neglected. 
%

\section{Charged particle motion}
\begin{figure*}[t!]
\label{fig_circular_orbits}
\centering
\includegraphics[width=.49\linewidth]{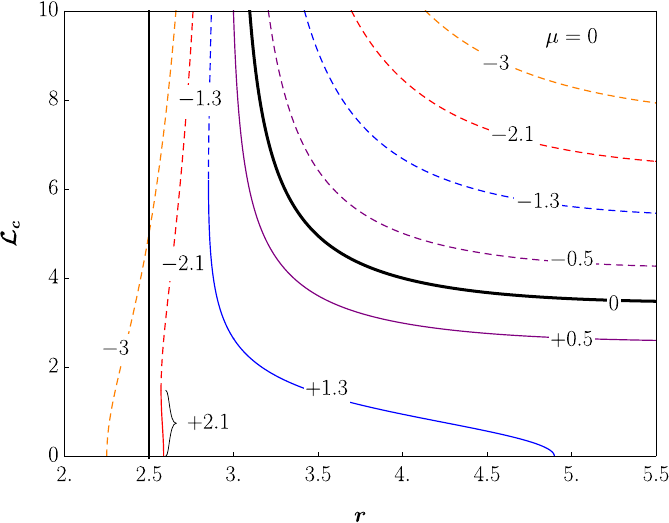}
\includegraphics[width=.49\linewidth]{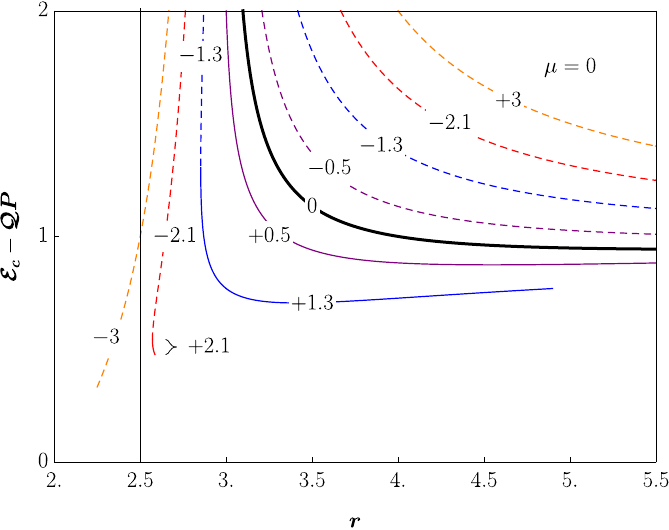}
\includegraphics[width=.49\linewidth]{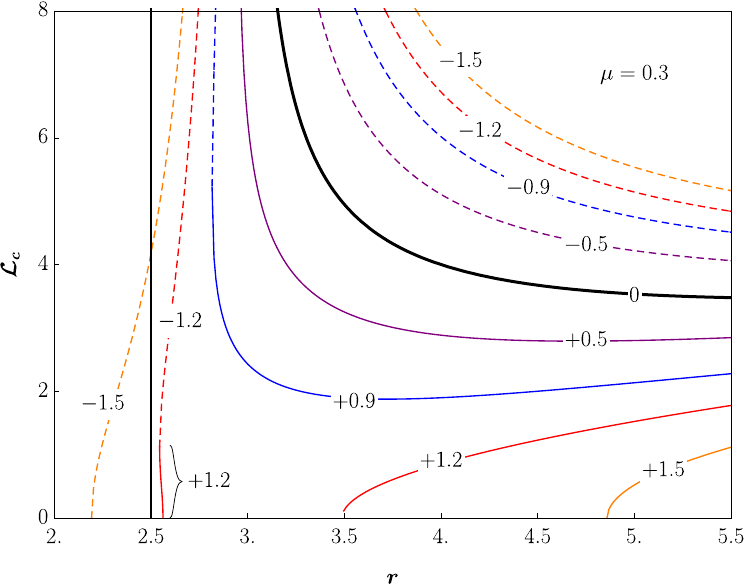}
\includegraphics[width=.49\linewidth]{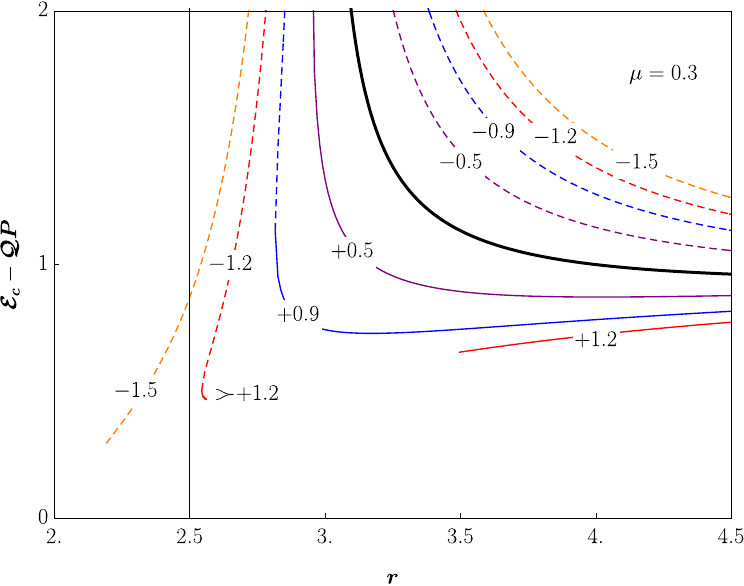}
\includegraphics[width=.49\linewidth]{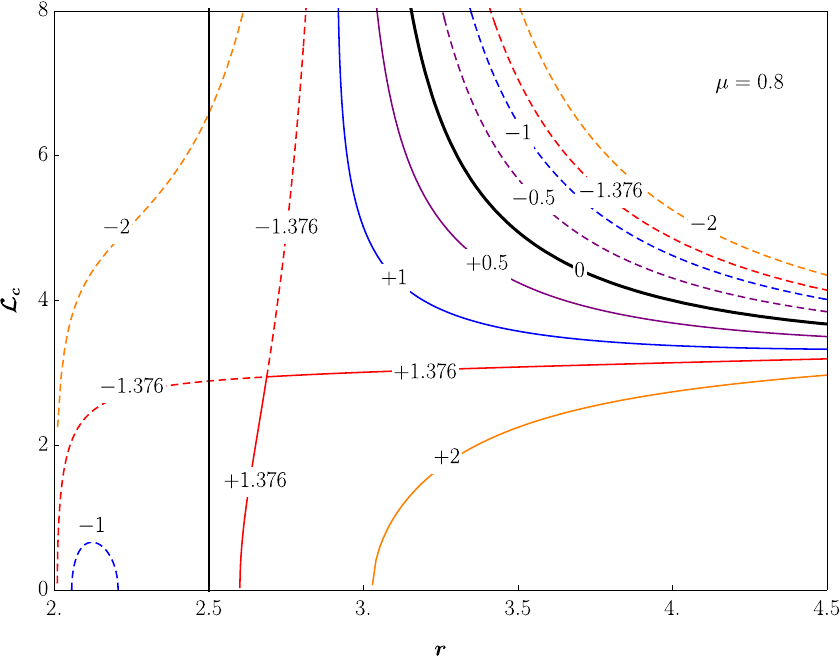}
\includegraphics[width=.49\linewidth]{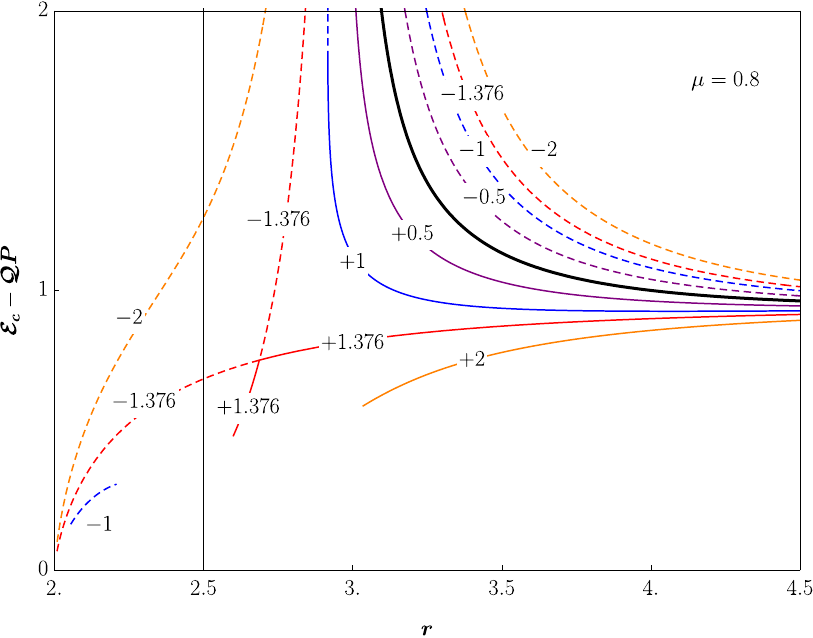}    
\caption{Behaviour of the specific angular momentum, $\mathcal{L}_{\rm c}$, and the redshift factor, $-u_{t}=\mathcal{E}_{\rm c}+q_{\rm{s}}A_{t}=\mathcal{E}_{\rm c}-\mathcal{Q}P$ of the circling charged particle. The numbers above the curves indicate the charge parameter $\mathcal{Q}$ values; the dashed curves correspond to $\mathcal{Q}<0$ cases, while the solid curves to $\mathcal{Q}>0$ cases.}
\label{E_c_L_c_rlsco}
\end{figure*}
\begin{figure*}[t!]
\centering
\includegraphics[width=.49\linewidth]{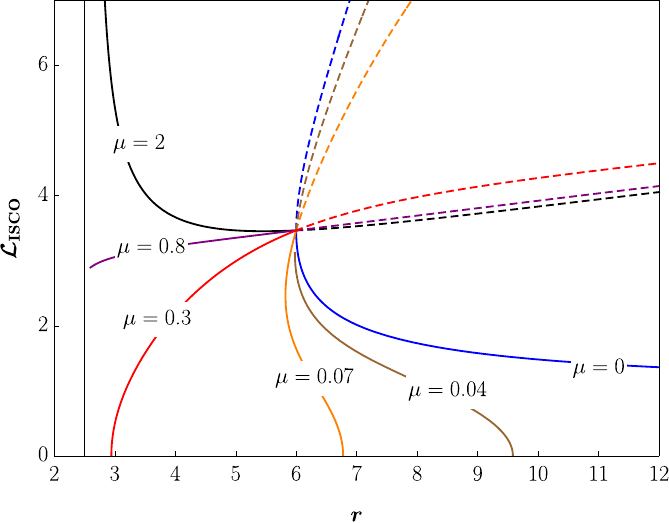}
\includegraphics[width=.49\linewidth]{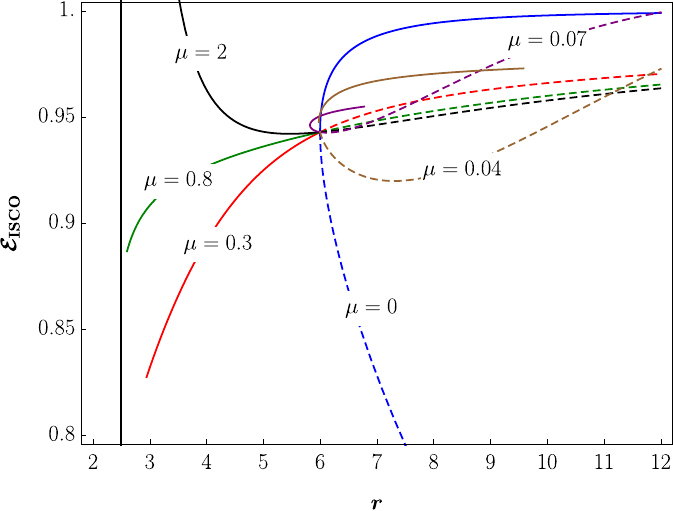}   
\caption{Behavior of the functions $\mathcal{L}_{\rm{ISCO}}$ and $\mathcal{E}_{\rm{ISCO}}$.}
\label{E_isco_L_isco_Risco}
\end{figure*}
\begin{figure*}[t!]
\centering
\includegraphics[width=.48\linewidth]{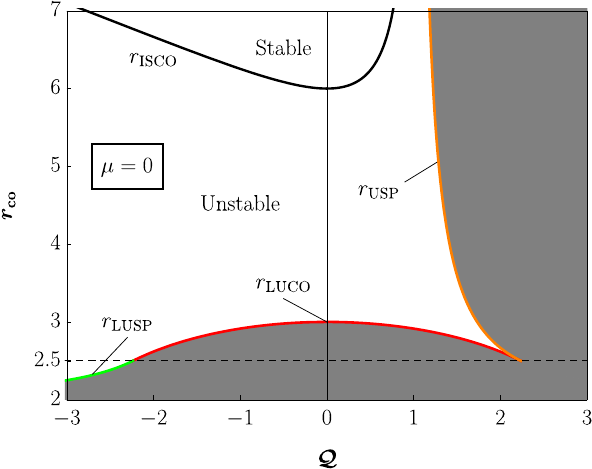}
\includegraphics[width=.48\linewidth]{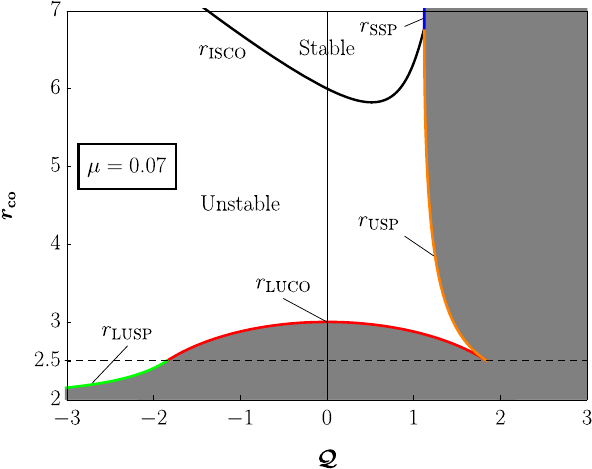}
\includegraphics[width=.48\linewidth]{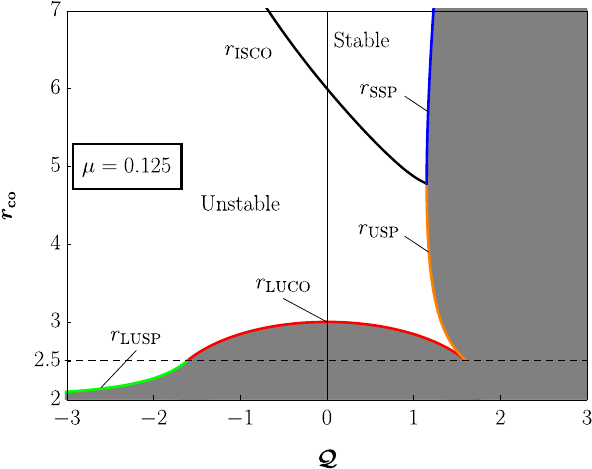}
\includegraphics[width=.48\linewidth]{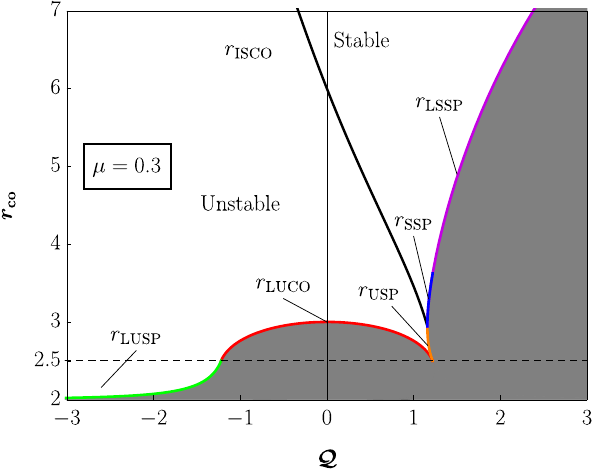}
\includegraphics[width=.48\linewidth]{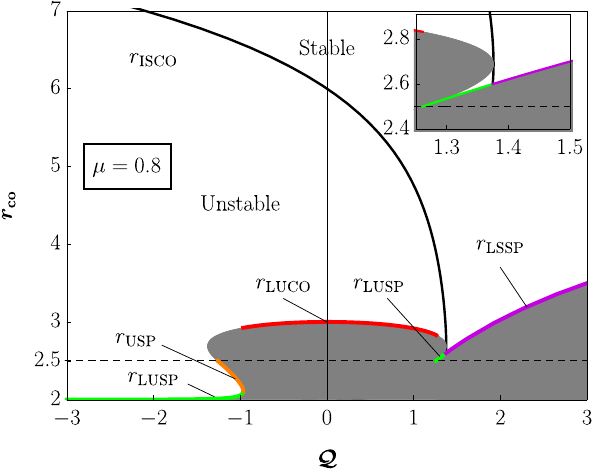}
\includegraphics[width=.48\linewidth]{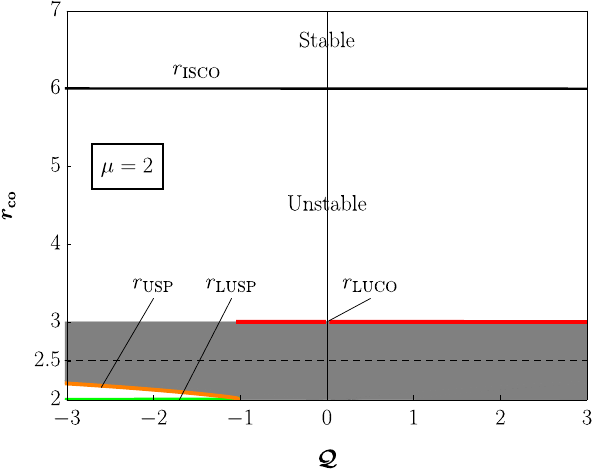} 
\caption{Locations of radii  $r_{\rm{ISCO}}$ (black curve), $r_{\rm{LUCO}}$ (red curve) $r_{\rm{SSP}}$ (blue curve), $r_{\rm{USP}}$ (orange curve), $r_{\rm{LSSP}}$ (purple curve) and $r_{\rm{LUSP}}$ (green curve) depending on the charge interaction parameter $\mathcal{Q}$ for different values of $\mu$. In the shaded area, there are no circular orbits of charged massive particles.}
\label{rlsco_Q}
\end{figure*}
%
\subsection{Angular velocity}
In this section, we study the motion of massive charged particles, assuming the most general situation of charges being the Proca ones. The limits of Maxwellian electric charges can be obtained by corresponding simplifications of more general Proca equations. The motion of massive photons are described in the next section. Similar studies of charged particle motion and related effects in the presence of magnetic field have been studied in detains in the work \cite{2020Univ....6...26S}. 

Hereafter, we omit the tilde marks everywhere for brevity and consider the case where a test particle with charge $q$, angular momentum $L$, and mass $m$ moves around a black hole with a Proca charge $Q$. The motion of the particle is governed by the Lagrangian density
\begin{equation}
    \mathcal{L}_{\text{den}}=\frac{1}{2}g_{\alpha\beta}\dot{x}^{\alpha}\dot{x}^{\beta}+\frac{q}{m}A_{\alpha}\dot{x}^{\alpha} .
\end{equation}
From Euler-Lagrange equation we get the equation of motion in the form
\begin{equation}
    \frac{{\rm d}u^{\mu}}{{\rm d}\tau}+\Gamma^{\mu}_{\alpha\beta}u^{\alpha}u^{\beta}-\frac{q}{m}F^{\mu}_{\hspace{1.75mm}\nu}u^{\nu}=0,
    \label{geodesic_eq}
\end{equation}
In case of a spherically symmetric metric with a vector potential \eqref{vec_pot_eq}, it is easy to consider the case when the particles are distributed in the equatorial plane, $\theta=\pi/2$, with the four-velocity $u^{\alpha}=\left(u^{t},0,0,u^{\phi}\right)$. In this case, assuming the normalization condition, $u^{\alpha}u_{\alpha}=-1$, the radial equation \eqref{geodesic_eq} yields the system of equations
\begin{eqnarray}
    -\frac{(u^{t})^{2}}{r^2}+r(u^{\phi})^{2}-q_{s}A_{t}'u^{t}&=&0,\\
    -\left(1-\frac{2}{r}\right)(u^{t})^{2}+r^{2}(u^{\phi})^{2}&=&-1,
\end{eqnarray}
where $q_{s}=q/m$ is a specific charge, and prime denotes the derivative with respect to the radial coordinate. The solution of these equations then reads 
\begin{eqnarray} \label{equphi}
u^{\phi}_{\pm}&=&\sqrt{\frac{1+(r-2)u^{t}_{\pm}r q_{s}A_{t}'}{(r-3)r^2}}, 
\label{phi_vel}
\\    u^{t}_{\pm}&=&\frac{(r q_{s}A_{t}')^{2}\pm\sqrt{4r(r-3)+r^2(r q_{s}A_{t}')^{2}}}{2(r-3)}. \label{equt}
\end{eqnarray} 
A similar approach has been used in the work \cite{2010CQGra..27d5001B}, where the authors has considered pure magnetic field acting on charged particle. 

The angular velocity with respect to distant observers, i.e. the coordinate angular velocity, is defined by the relation 
\begin{eqnarray}
\Omega_{\pm}=u^{\phi}_{\pm}/u^{t}_{\pm},
\end{eqnarray}
giving the Keplerian velocity, $\Omega_{\pm}=\pm 1/r^{3/2}$, in the case of $A_{t}=\text{const}$ or $q_s=0$. Note that the full explicit form of the angular velocity $\Omega_{\pm}$ involving terms (\ref{equphi}) and (\ref{equt}) is rather complex, and it is much more convenient to express it in terms of the specific angular momentum defined below.\\

\subsection{Effective potential and existence of orbits}
A particle moving in the equatorial plane of a static and spherically symmetric spacetime is subject to the conservation of its specific energy and angular momentum
\begin{eqnarray}
    -\mathcal{E}&=&-E/m=-\left(\frac{r-2}{r}\dot{t}-q_{s}A_{t}\right), \\
\mathcal{L}&=&L/m=r^2\sin^2{\theta}\dot{\phi}. 
\end{eqnarray}
Then, the radial equation of motion \eqref{geodesic_eq} can be written in the simple form
\begin{equation}
\dot{r}^2=\left(\mathcal{E}-V_{+}\right)\left(\mathcal{E}-V_{-}\right),
\end{equation}
where
\begin{equation} 
V_{\pm}=\frac{\mathcal{Q}e^{-\mu(r-2)}}{r}U[\mu,0,2\mu(r-2)]\pm\sqrt{\left(1+\frac{\mathcal{L}^2}{r^2}\right)\left(1-\frac{2}{r}\right)}\nonumber
\end{equation}
stands for the so-called effective potential of the charged test particle, $\mathcal{Q}=q_{s}Q$ represents the parameter describing the non-gravitational interaction between the particle and Proca black hole (interaction parameter), and the dot denotes the derivative with respect to the proper time $\tau$.
In the following, we consider the $V_{+}$ solution, corresponding to the so-called positive root states with the future oriented time \cite{1973grav.book.....M,1989BAICz..40...65B,1989BAICz..40..133B}. Therefore locations of circular orbits can be determined by solutions of the system of equations $\mathcal{E}=V_{+}$ and $\mathrm{d}V_{+}/\mathrm{d}r=0$ that can be written in the form  
\begin{eqnarray}  
\mathcal{Q}P(r)+J(r)&=&\mathcal{E},\\
    -\mathcal{L}^2\left(r-3\right)+r^2+\mathcal{Q}r^4J(r)P'(r)&=&0\label{firs_der}, 
\end{eqnarray}
where 
\begin{eqnarray}
    P(r)&=&\frac{e^{-\mu(r-2)}}{r}U[\mu,0,2\mu(r-2)],\\
    J(r)&=&\sqrt{\left(1-\frac{2}{r}\right)\left(1+\frac{\mathcal{L}^2}{r^2}\right)}.  
\end{eqnarray}

From equation \eqref{firs_der} we find the angular momentum of a circling particle in the form 
\begin{align}
\label{L_circ}
\mathcal{L}_{c} 
  &= \pm \Biggl(
     \frac{r^{2}}{r - 3}
     + \frac{(r - 2)\, r^{7/2} P'}{2(r - 3)^{2}}
       \Bigl(
         \mathcal{Q}^{2} r^{3/2} P' \notag \\
  &\qquad\qquad
         + \mathcal{Q} 
           \Bigl[\, 4(r - 3) + \mathcal{Q}^{2} r^{3} P'^{2} \Bigr]^{1/2}
       \Bigr)
     \Biggr)^{1/2}
\end{align}
where $\pm$ indicates the direction of the particle circulation. Due to the considered spherical symmetry, however, we can only assume the positive direction of the circulation. Note that the first term under the square root corresponds to the well-known angular momentum of particle circling in the background of the Schwarzschild black hole.

As we can see from relation (\ref{L_circ}), the circular orbits can only exist in the regions where   
\begin{equation}
\mathcal{L}^2_{c}\geq 0, \quad 4(r-3)+\mathcal{Q}^2r^3P'^2\geq 0.
\label{ineq_1}
\end{equation}
The boundaries of these regions (boundary curves) are then determined by relations   
\begin{eqnarray}   
\label{LUCO_curve}
\label{SP_curve_0}
    \mathcal{Q}&=&\pm\frac{2\sqrt{3-r}}{P'r^{3/2}}, \\
    \label{SP_curve}
    \mathcal{Q}&=&\pm\frac{1}{|P'|r^{3/2}\sqrt{r-2}}, 
\end{eqnarray}

It is interesting to note that the intersection of these curves is located at $r=5/2$. This feature holds for all ranges of parameters regardless of the type of potential, which is simply a result of the solution of the equation $2\sqrt{r-2}\sqrt{3-r}=1$. Despite the fact that the surface defined by equations \eqref{SP_curve_0} and \eqref{SP_curve} limits only the angular momentum of the circling particles, $\mathcal{L}_{\rm c}$, it should be emphasized that the same holds for their energies, $\mathcal{E_{\rm{c}}}$ even if the square of angular momentum appears in the definition of energies. For different values of the charge interaction parameter, $\mathcal{Q}$, and the Proca parameter, $\mu$, the behaviour of the angular momentum and the redshift factor are shown in Fig.~\ref{E_c_L_c_rlsco}. The important features here is the limitation of the circular orbits existence by the radius $r=5/2$ supposed the  black hole and a test particle have the same charge signs (see Fig.~\ref{E_c_L_c_rlsco}).

The position of the innermost stable circular orbit (ISCO), due to the condition $V''=0$, is determined by the equation 
\begin{align*}
     &(3-2r)r^4+\mathcal{L}_{\rm c}^4(15+2(r-6)r)+\\
     &\mathcal{L}_{\rm c}^2r^2(22+3(r-6)r)-
     r^4J(r)^2(r^2+\mathcal{L}_{\rm c}^2(r-3))\frac{P''}{P'}=0,
\end{align*}
and the related energy and angular momentum of a charged particle at ISCO are plotted in Fig.~\ref{E_isco_L_isco_Risco}. There are only two independent solutions of this biquadrate equation, $\mathcal{L}^2_{\pm}$, satisfying the condition $\mathcal{L}^2_{\pm}\geqslant0$. 
The position of the ISCO depending on the charge parameter is shown in Fig. \ref{rlsco_Q}. By studying circular orbits, it is also necessary to consider the last unstable circular orbit (LUCO) that is, due to the condition, \eqref{ineq_1}, determined by relation 
\begin{eqnarray*}
   r_{\rm LUCO}=\frac{1}{2}\left(3+\sqrt{9-\mathcal{Q}^2}\right),\quad -\sqrt{5}\leq\mathcal{Q}\leq\sqrt{5},
\end{eqnarray*}

below which no circular orbit exists and all trajectories are directed to the event horizon or to infinity.  
We can see that stable orbits can also exist for values  $\mathcal{Q}\geq 1$, which is not possible in the case $\mu=0$, when all stable orbits are located in the range $-\infty<\mathcal{Q}< 1$ \cite{Tursunov:2021jjf}.

Among interesting properties of weakly charged \\ Schwarzschild black holes is the existence of stationary radii (points) determined by the condition $\mathcal{L}_{\rm c}=0$. At these points, the gravitational attraction is fully compensated by an electromagnetic repulsion. In the weakly Proca-charged black hole case, we can identify two types of stationary points: stable stationary points (SSP) and unstable stationary points (USP) (see Fig.~\ref{rlsco_Q}).
If there are no circular orbits below stable and unstable stationary points, we denote these points as LSSP and LUSP, respectively. In the case of using the Coulomb potential, USP and LUSP points are located at the radii 
\begin{eqnarray*}
     r_{\rm USP}&=&\frac{2\mathcal{Q}^2}{\mathcal{Q}^2-1},\quad 1<\mathcal{Q}< \sqrt{5},\\
     r_{\rm LUSP}&=&\frac{2\mathcal{Q}^2}{\mathcal{Q}^2-1},\quad -\infty<\mathcal{Q}< -\sqrt{5},
\end{eqnarray*}
and SSP and LSSP exist only for $\mu\neq 0$ (see also Fig. \ref{rlsco_Q} with $\mu=0$). 
It must be pointed out that LUSP is not physically realistic, since no stationary points can exist in the attractive Coulomb potential case. We, however, consider this situation for completeness. The boundary curve \eqref{SP_curve} describes all stable and unstable stationary points, while the boundary curve \eqref{LUCO_curve} describes the last unstable circular orbit. 

Positions of ISCO changes dramatically for non-zero values of the Proca parameter. For instance, for $\mathcal{Q}>0$ and $\mu\geq 0.125$, the radius of ISCO decreases with increasing $\mathcal{Q}$, while for $\mu=0$ it increases with $\mathcal{Q}$ and diverges at $\mathcal{Q}=1$. Furthermore, while in the Coulomb potential case with $\mu=0$ all stable/unstable circular orbits exist within the range $-\infty < \mathcal{Q}\leq\sqrt{5}$, for $\mu\neq$ no such boundary has been found. Therefore the circular orbits can be present for all values of the charge parameter $\mathcal{Q}$. We can also see that for large values of $\mathcal{Q}>0$ all stable orbits end at stationary points, and that for large values of the Proca parameter $\mu$ the ISCO is essentially independent of the charge parameter and located at a distance of $r=6$. (see Fig. \ref{rlsco_Q}). For $\mu \gg 1$, the radius  of ISCO can be approximately determined by the relation 
\begin{equation}
    r_{\rm{ISCO}}=6-\frac{9+225\mu}{2^{\frac{1}{2}+3\mu}e^{4\mu}\mu^{\mu-1}}\mathcal{Q},
\end{equation}
with the linear dependence on the charge parameter $\mathcal{Q}$. 
%
 
\section{Observational properties of weakly Proca-charged black hole}
In this section we discuss possible observational properties of the weakly Proca-charged black hole by considering two independent gravitational tests: gravitational lensing leading to the black hole shadow, and the fitting of orbital parameters to data from measurements of the hot-spots motion around the Galactic center of the supermassive black hole Sgr~A*. 

\subsection{Gravitation bending of massive photons and black hole shadow}
\begin{figure*}[t]
\centering
\includegraphics[width=12.cm]{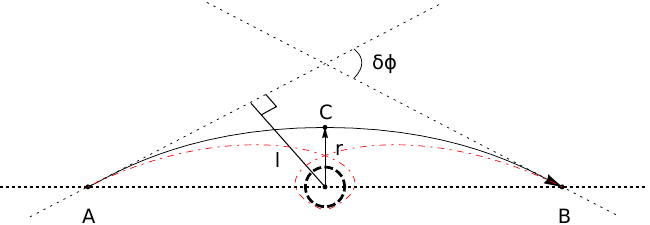}
\caption{\label{fig:epsart} Schematic representation of gravitational lensing effect. The motion of a photon from point $A$ to $B$ is presented in two ways. The black curve represents motion at $\phi_{tot}=\pi$ and red when it makes one revolution around a black hole $\phi_{tot}=3\pi$. The last unstable photon circular orbit is  depicted by solid dashed curve}.
\end{figure*}
\begin{figure*}[t]
\centering
\includegraphics[width=.48\linewidth]{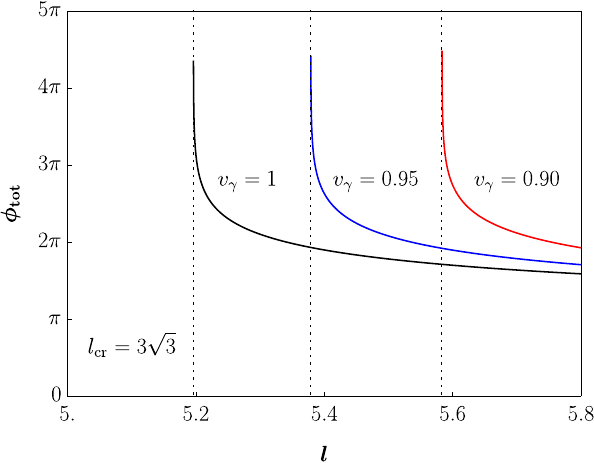}
\includegraphics[width=.48\linewidth]{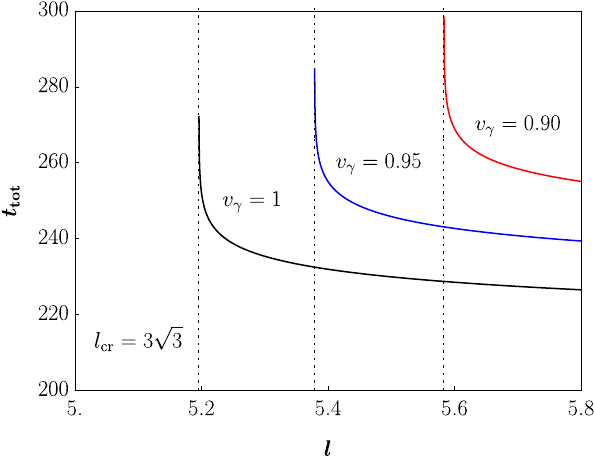}
\includegraphics[width=.48\linewidth]{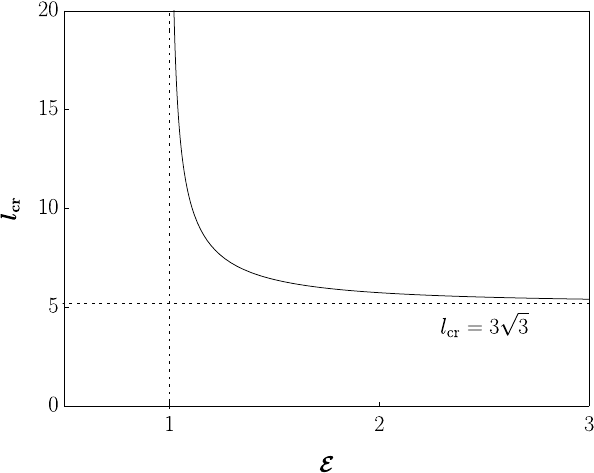}
\includegraphics[width=.48\linewidth]{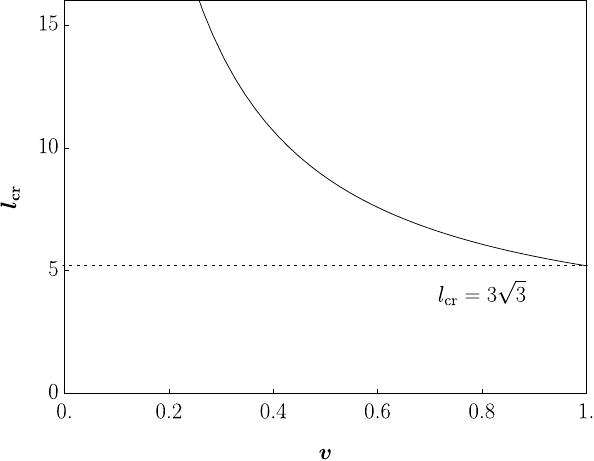}
\caption{Upper row: deflection angle, $\phi_{\rm tot}$, and total time dilation, $t_{\rm tot}$, in dependence on the impact parameter, $l$. Lower row: the critical impact parameter, $l_{\rm cr}$, versus energy of massive photon, $\mathcal{E}$ (lower left) and critical impact parameter versus photon's velocity (lower right).}
\label{Shadow_E_L}
\end{figure*}
In this section, we present a gravitational bending of massive photons by discussing the possibility of construction of the black hole shadow created by these photons. From the normalization condition and conservation of angular momentum, we find for the radial and angular motion of an uncharged particle
\begin{eqnarray}   
(\dot{r})^2&=&\mathcal{E}^2-f\left(1+\frac{\mathcal{L}^2}{r^2}\right) , 
    \label{eq_rdot}
    \\(\dot{\phi})^2&=&\frac{\mathcal{L}^2}{r^4}. 
    \label{eq_phidot}
\end{eqnarray}
A schematic representation of the gravitational lensing effect in the equatorial plane is shown in Fig.\ref{fig:epsart}. Here an electromagnetic signal originates from the point $A$ with coordinates $(r_{\rm A},\phi_{\rm A})$ and reaches the point $B$ with coordinates $(r_{\rm B},\phi_{\rm B})$, passing by the central black hole at the nearest point $C$ without being captured. Then from relation \eqref{eq_rdot} we obtain
\begin{equation}
{\rm d}t=\frac{{\rm d}r}{f\sqrt{1-f\left[1+v_{0}^2\left(\frac{l^2}{r^2}-1\right)\right]}},
\end{equation}
where we used $\dot{r}=({\rm d}r/{\rm d}t)({\rm d}t/{\rm d}\tau)=({\rm d}r/{\rm d}t)\dot{t}$
and rewrote the energy and angular momentum in terms of a photon velocity $v_0$, namely, $\mathcal{E}=1/\sqrt{1-v_{0}^2}$ and $\mathcal{L}=lv_{0}/\sqrt{1-v_{0}^2}$. 

The impact parameter, $l$ can be then determined from the relation $lv_{0}={\mathcal{L}}/{\mathcal{E}}$, which is consistent with the definition $l={\mathcal{L}}/{\mathcal{E}}$ in the case of $v_{0}=1$ for the zero mass photon \cite{2016CQGra..33q5014L,2020EPJC...80..835H}. By using the condition $\dot{r}=0$ we can find the impact factor in terms of the  closest distance $r_{\rm C}$
\begin{equation}
    l^2=\frac{r^2_{\text{C}}}{f(r_{\text{\rm C}})}W(r_{C}) ,
\end{equation}
where 
\begin{equation}
    W(r_{\rm C})=f(r_{\rm C})+\frac{2}{r_{\rm C}v^2_{0}},\quad f(r_{\rm C})=1-\frac{2}{r_{\rm C}},
\end{equation}
which, in the case of the mass-less photon, $v_{0}=1$, gives $W(r_{\rm C})=1$. Therefore, we find  
\begin{equation}
     {\rm d}\phi=\frac{l{\rm d}r}{r^2\sqrt{W(r)-\frac{l^2f}{r^2}}}.
\end{equation}
for the deflection angle, and consequently  
\begin{eqnarray} 
\phi_{\text{tot}}&=&2\int_{r_{C}}^{r_{0}}=\frac{l{\rm d}r}{r^2\sqrt{W(r)-\frac{l^2f}{r^2}}},\\ 
t_{\text{tot}}&=&2\int_{r_{C}}^{r_{0}}\frac{{\rm d}r}{v_{0}f\sqrt{W(r)-\frac{l^2f}{r^2}}}, 
\end{eqnarray} 
where $t_{\text{tot}}$ is the total travel time of the photon between the points $(r_{\rm A},\phi_{\rm A})=(r_{0},0)$ and $(r_{\rm C},\phi_{\rm C})$,  as considered in the paper \cite{2019EPJC...79...44S}, and the impact parameter, $l$, was chosen so that the photon reaches the point $(r_{B},\phi_{B})=(r_{0},\pi)$ after passing the point $(r_{\rm C},\phi_{\rm C})$.

If $\phi_{\text{tot}}=3\pi$, the photon completes one revolution around the black hole before reaching the point B. The impact factors for this motion are listed in Table \ref{tab_1}. To illustrate the time delays of photons with different velocities, we consider the emitter and detector at the distance of $r_{0}=100$ from the black hole. We can see that increasing the photon mass leads to the increase of the impact parameter and of the time delay of detected photons with different energies. Moreover, we can see that the time dilation $\Delta t=t_{\rm{tot}}-t_{\rm{tot}}^{*}$ for a photon finishing half of a revolution $\phi_{\rm{tot}}=\pi$ around the black hole is almost the same as for a photon finishing the full revolution, $\phi_{\rm{tot}}=3\pi$. Note that in the case of flat spacetime, the total travel time can simply be expressed as 
\begin{equation}    t_{\text{tot}}^{*}=\frac{2}{v_{0}}\sqrt{r_{0}^2-r_{C}^2}.
\end{equation}

\begin{table}[h]
\caption{\label{tab_1}%
A light deflection angle, total time and time dilation relative to the non-massive photon $\Delta t$ for various $v_0$.}
\centering
\begin{tabular}{ccccc}
\hline
\textrm{$v_{0}$} &
\textrm{$\phi_{\text{tot}}$} &
\textrm{$t_{\text{tot}}$} &
\textrm{$l$} &
\textrm{$\Delta t$} \\
\hline
$1.00$ & $\pi$   & $211.423$ & $15.747$ & $-$ \\
$0.95$ & $\pi$   & $221.909$ & $16.178$ & $10.486$ \\
$0.90$ & $\pi$   & $233.472$ & $16.665$ & $22.049$ \\
\hline
$1.00$ & $3\pi$  & $250.153$ & $5.202$  & $-$ \\
$0.95$ & $3\pi$  & $261.243$ & $5.385$  & $11.09$ \\
$0.90$ & $3\pi$  & $273.421$ & $5.590$  & $23.268$ \\
\hline
\end{tabular}
\end{table}
In the following we determine the shadow of a black hole that can be found from the photon equation of motion taking into account the fact that the photon motion is allowed only in the region where $\mathcal{E}^2\geq V^2$. By using the system of equations  
\begin{equation}  
V_{\rm{eff}}^2=\mathcal{E}^2,\quad V_{\rm{eff}}'=0, 
\end{equation}
where the effective potential is defined by relation \eqref{eq_rdot}, 
the condition $\mathcal{E}^2>V^2$ can be written in the form 
 \begin{equation}	l^2<l_{\rm{cr}}^2=\frac{r^2}{f}W(r) .
  \label{con_l}
 \end{equation}
 Therefore, the critical impact parameter $l_{\rm{cr}}$ can be expressed in the form
 \begin{equation}   
 l_{\rm{cr}}=\frac{\sqrt{-1+20v_{0}^2+8v_{0}^4+(1+8v_{0}^2)^{3/2}}}{\sqrt{2}v_{0}^2}, 
 \end{equation}
 being dependent on the initial velocity (see Fig. \ref{Shadow_E_L}); we can see that for $v_0=1$ we obtain the value $l_{\rm cr}=3\sqrt{3}$.  
Moreover, since the mass, wavelength, and velocity of a massive photon are related as~\cite{Robles:2012zz}
\begin{equation}
    v_{0}=\sqrt{1-\left(\frac{\mu\lambda}{2\pi}\right)^2}\approx1-\frac{1}{2}\left(\frac{\mu\lambda}{2\pi}\right)^2,
\end{equation}
the critical impact parameter determining the radius of the shadow of a black hole takes the simple form 
\begin{equation} 
\label{eq:rshadow}
l_{\rm{cr}}=3\sqrt{3}+\sqrt{3}\left(\frac{\mu\lambda}{2\pi}\right)^2, 
\end{equation}
 where the second (mass-wavelength) term describes the contribution of the photon mass and its wavelength. 

However, it is necessary to point out that the difference between the shadow radii in the case of massless and massive photons (the mass-wavelength term) is negligible for the wavelength range commonly used in radio astronomy. 
Suppose that the mass of a photon is equal to the upper limit determined by experiment, namely $m_{\gamma}\leq 10^{-47}{\rm g}$. Then, according to the observations of the Event Horizon Telescope (EHT), observing supermassive black holes M87 and Sgr~A* at the wavelength of $\lambda=1.3 \text{mm}$, one can estimate the mass-wavelength term value 
\begin{equation}
    \left(\frac{\mu \lambda}{2\pi}\right)^2\approx10^{-25}.
\end{equation}
Therefore, the influence of the photon mass cannot be tested using solely observations of the black hole shadows. 
Even for longer radio wavelengths, e.g. around $10^{5}\text{cm}$,  known as a "VLF" (very low frequency) signal, also commonly used for submarine communications or for Earth ionosphere investigation, the mass-wavelength term takes the value  
\begin{equation*}
    \left(\frac{\mu \lambda}{2\pi}\right)^2\approx 10^{-13}.
\end{equation*} 
Consequently, by looking at the black hole shadow, only extremely cold photons could demonstrate the existence of non-zero photon mass. However, such photons would easily be scattered before reaching the distant observer. 

\subsection{Fitting the period-radius orbits of the Galactic center flares}
\begin{figure*}[t]
\centering
\includegraphics[width=8.cm]{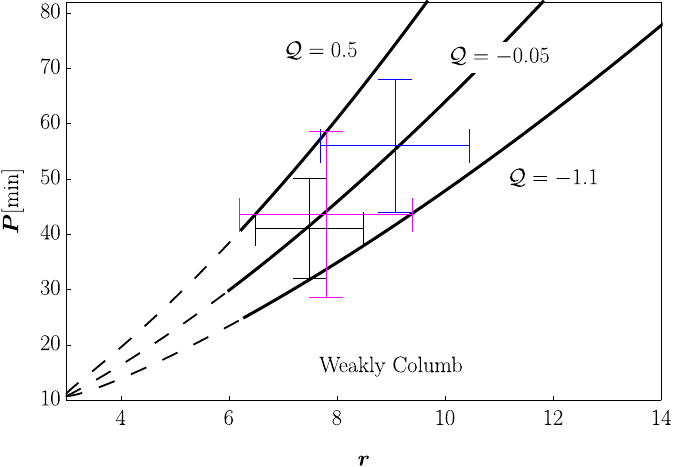}
\includegraphics[width=8.cm]{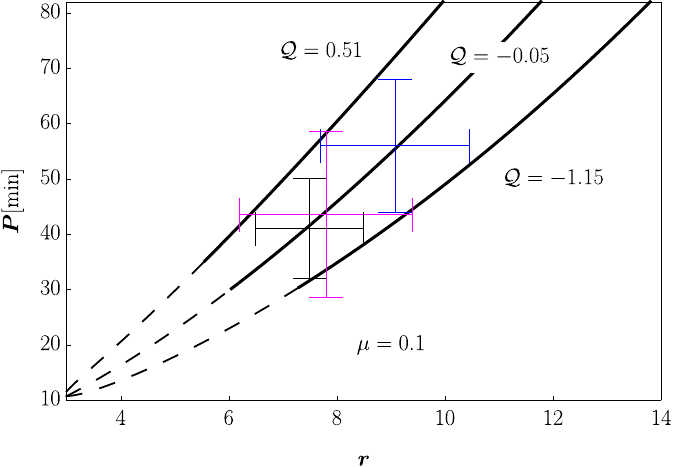}
\includegraphics[width=8.cm]{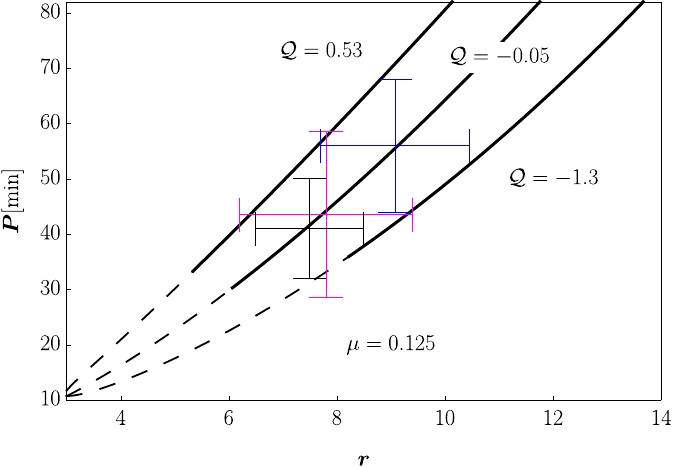}
\includegraphics[width=8.cm]{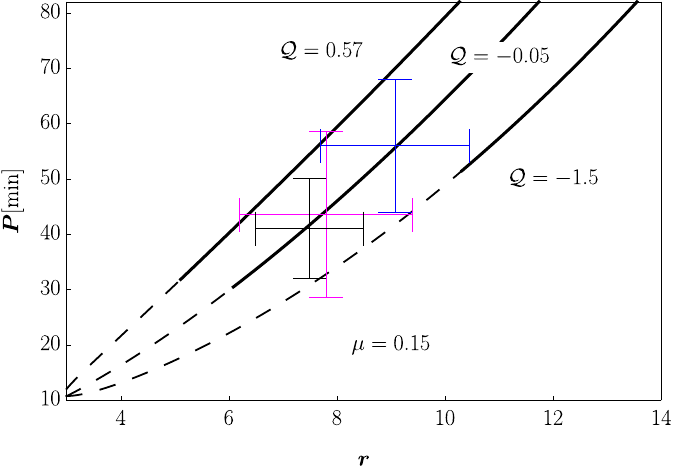}
\caption{Orbital period–radius relations of three flares observed by GRAVITY in 2018 on July 22 (black), 
May 27 (pink), and July 28 (blue) fitted with circular orbits of a charged hot spot moving around a Schwarzschild black hole of mass $M=4\times 10^{6}{\rm M}_{\odot}$ carrying a small Proca charge, $Q$. The dynamics of the hot spots is characterized by the interaction parameter $\mathcal{Q}=q_{\rm hs} Q$, where $q_{\rm hs}$ denotes the hot spot specific charge, reflecting the Proca interaction between the hot spot and the black hole. The dashed  curves indicate unstable circular motion and the bold curve represents a stable orbit.}
\label{Flare_fig}
\end{figure*}
In this section, we test the potential role of the Proca charge of the black hole in fitting the period-radius relations of the three bright flares observed in the vicinity of the supermassive black hole Sgr~A* at the Galactic centre by the near-infrared GRAVITY instrument \cite{2018A&A...618L..10G}. The orbital motion of the flares has been detected at a distance of about $6$ to $11$ gravitational radii ($GM/c^2$), given a black hole mass of about $4 \times 10^6 M_{\odot}$. The proximity of the orbits to the black hole and the relativistic velocities of the flares provide an opportunity to test gravity in a strong regime. Similar tests in general relativity and alternative gravity theories have recently been studied in the paper \cite{2022EPJC...82..407S}.    

A plasma moving relativistically around a black hole at the Galactic centre undergoes charge separation due to an ordered magnetic field detected in the vicinity of the black hole \cite{2020ApJ...897...99T}. This charge arises as a result of the compensation of the comoving frame electric field for the motion of the plasma in the external magnetic field. It has been shown that the net charge of the flare plasma, although tiny, can be dynamically significant. Therefore, it can potentially interact electrostatically with the black hole if the Proca or electric charge is present at its centre. Assuming that the flare components are on circular orbits, one can calculate the orbital period in terms of the Keplerian frequency $\Omega$ as follows 
\begin{equation}
    P=\frac{2\pi}{6
    0}\left(\frac{GM}{c^3}\right)\frac{1}{\Omega} {\rm min}, 
\end{equation}
where $\Omega$ in a weakly Proca-charged black hole case is a function of a distance and interaction parameter, $\mathcal{Q}$
\begin{equation}
\Omega = \frac{2(r-3)\mathcal{L}_{\rm c}}{\mathcal{Q}r^4P'+r^2\sqrt{4(r-3)r+\mathcal{Q}^2r^4P'^2}},
\end{equation}
where $\mathcal{L}_{\rm c}$ defined by relation \eqref{L_circ}. For negative values of the charge parameter, we have a discontinuity point at $r=3$. Therefore, the behavior of the angular velocity at $\mathcal{Q}<0$ can be divided into two parts, where up to $r<3$ the angular velocity is negative and at $r>3$ it is a positively defined function.

In Fig.~\ref{Flare_fig} we fit the period-radius relations of three flares observed in 2018 on July 22, May 27, and July 28 at the Galactic center with the model of a non-rotating weakly charged Proca black hole. As a special case of the Proca theory, we include the case of the weakly electrically charged Reissner-Nordstr\"{o}m black hole. The dashed parts of the curves represent regions below the ISCO position, while the solid parts of the curves show the behavior above the ISCO. 
We can see that in the case of absence of the Proca parameter, $\mu=0$, the positive/negative branch of the interaction parameter $\mathcal{Q}$ can fit the observations in terms of stable circular orbits. The inclusion of a small, but non-zero Proca parameter, $\mu\neq0$, destabilises the orbits for $\mathcal{Q}<0$. Therefore, for relatively large values of $\mu$, the orbits with $\mathcal{Q}<0$ become unstable again at the given points of the $P-R$ diagram. The value of the interaction parameter then lies in the range $-1.1 < \mathcal{Q} < 0.5$.  Assuming that all orbits of the flares are stable within the error bars of the measurements, then we obtain the new constraint on the Proca parameter that gives the range $0\leq\mu\leq 0.125$. 

\section{Discussion \& conclusion}
We have considered a special case of the Einstein-Proca theory in a weak field approximation.  
 As expected, in this regime the background geometry is described by the Schwarzschild metric, whereas the vector potential acquires a non-trivial modification. We have derived its exact form, analysed its asymptotic behaviour, and shown that it decays more rapidly than the Yukawa potential typically associated with massive vector fields in flat spacetime. Importantly, when extending the analysis to higher-order terms in the expansion, we find that these contributions diverge only in the immediate vicinity of the event horizon. This indicates that the weak-field approximation, while well-behaved at large radii and essentially up to the horizon, cannot be straightforwardly extrapolated all the way into the near-horizon region, where a fully non-perturbative treatment would be required. 
 
 The Proca parameter describing massive boson interactions also allows one to estimate a possible contribution of the photon mass to the vector potential in a non-trivial analytical form. We have parameterised the Proca-charge by a dimensionless parameter $\mu$, which after the recovery of the physical constants, takes the form 
\begin{equation*}
 \mu \equiv    \frac{m_{\gamma}c}{\hbar}\left(\frac{G M}{c^2}\right), 
\end{equation*}
which couples the photon mass $m_{\gamma}$ with that of the back hole $M_{\rm BH}$. 
If we assume an extremely tiny photon mass, being $m_{\gamma}\leq 10^{-47}\text{g}$, i.e. below the current experimental uppermost limit (based on the examination of Coulomb's law deviations on small scales), one can find that the Proca-charge parameter $\mu$ can be dynamically significant, i.e. being of the order of unity, if the black hole mass is above the following value 
\begin{equation*}
M\geq 10^{4}M_{\odot}.
\end{equation*}
This means that the effect of the Proca charge increases with the mass of the black hole, and for supermassive black holes it becomes much more significant than for those of stellar mass. This potentially offers an interesting way to tighten the current experimental constraints on the photon mass by the observations of supermassive black holes. Of course, such a constraint requires precise measurements of the black hole charge, since for weakly Proca-charged black holes the Proca parameter $\mu$ is not independent but appears to be coupled to the small central charge of the black hole.  There have already been attempts to measure the black hole charge based on various observational methods (see, e.g. \cite{2018MNRAS.480.4408Z}), with the most progress achieved in the case of the Galactic center black hole Sgr~A* \cite{2019JPhCS1258a2031Z,2019Obs...139..231Z}), concluding that the charge of the black hole at the Galactic center is most likely non-zero and positive. Using their upper limit on the charge $Q_{BH}\leq 10^{24}\text{Fr}$, and again the upper limit on the photon mass as $m_{\gamma}\leq 10^{-47}\text{g}$, one can obtain the deviation of the spacetime metric from that of the Schwarzschild spacetime as $\epsilon\leq 10^{-19}$. This implies that the effect of the Proca charge is unlikely to be distinguishable by gravity tests. However, the above estimate shows that even tiny photon mass, if present, can have considerable effect on the dynamical and electromagnetic environment of supermassive black holes.  

Next, we studied the motion of a charged particle around a weakly Proca-charged black hole. We  obtained a general expression for the angular velocity and angular momentum of the circular motion. It is known that in the case of the Coulomb potential, the ISCO exists in the range $\infty<\mathcal{Q}\leq 1$, although at $\mathcal{Q}=1$, the ISCO is located at infinity. However, in the Proca case with a non-zero parameter $\mu$, the ISCO may exist beyond $\mathcal{Q}>1$ and be finite for any charge parameter. Additionally, for some values of $\mu$, the ISCO can exist up to the critical point $r=2.5$, beyond which circular motion is not possible. The point $r=2.5$ represents the closest last unstable circular orbit for a positive charge parameter. It is universal and does not depend on the choice of potential; the only condition for the existence of this point is the existence of a vector potential. From the perspective of a test particle, the charged Proca black hole at large distances, $r-2> 1/\mu$, is almost identical to a neutral black hole. 

The motion of a massive photon around a black hole, in particular, the deflection angles and the time dilation depending on the velocities of photons have been calculated and analyzed. Although the deflection angles of a massive photon have been studied in detail by other authors, the shadow of the black hole has not been studied previously. We have shown that the presence of mass does indeed increase the shadow of a black hole, depending on the velocities or energy of the photon. However, in realistic situations, we have calculated that even low-energy photons would not be able to significantly increase the shadow of a black hole if they had a non-zero mass of about the current upper limit.

As a final test, we have discussed a possible role of the Proca-charge in the fitting of the orbits of three near-infrared flares observed by the GRAVITY around the Galactic center supermassive black hole. Although the error bars can be covered without the Proca parameter, the lower periods can only be fitted by unstable orbits. If we assume that all orbits of the flares within the error bars of the measurements are stable, then we obtain the constraint on the Proca parameter within the range of $0 \leq \mu\leq 0.125$.  However, to confirm these constraints, more observations of flare motion at the Galactic center by GRAVITY are needed, with the observations lasting 2–3 revolutions around SgrA*, i.e., 4–6 times longer than currently available. 
%
\section{Acknowledgments}
 
The authors acknowledge support from the Czech Science Foundation Grant No.~\mbox{23-07043S} and the Silesian University in Opava Grant No. SGS/24/2024. 
Abylaikhan Tlemissov acknowledges  the support of science and research in the Moravian-Silesian Region 
Grant No.  RRC/09/2023. 
Arman Tursunov thanks the Alexander von Humboldt Foundation for its Fellowship. 
\appendix
\begin{figure*}[t]
\centering
\includegraphics[width=.49\linewidth]{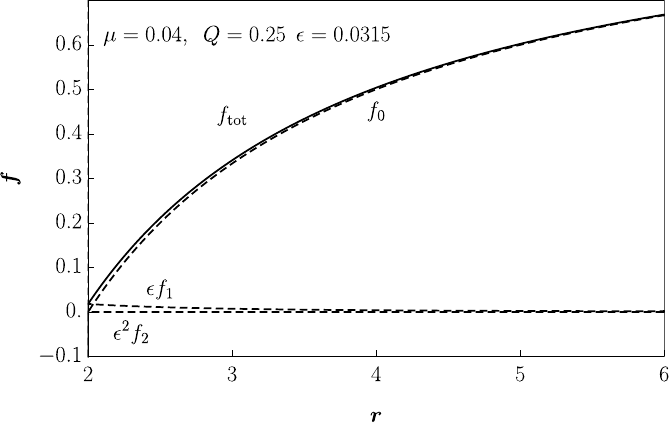}
\includegraphics[width=.49\linewidth]{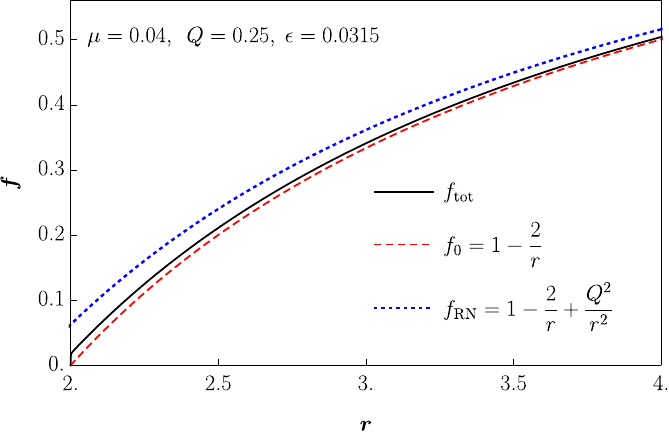}
\includegraphics[width=.49\linewidth]{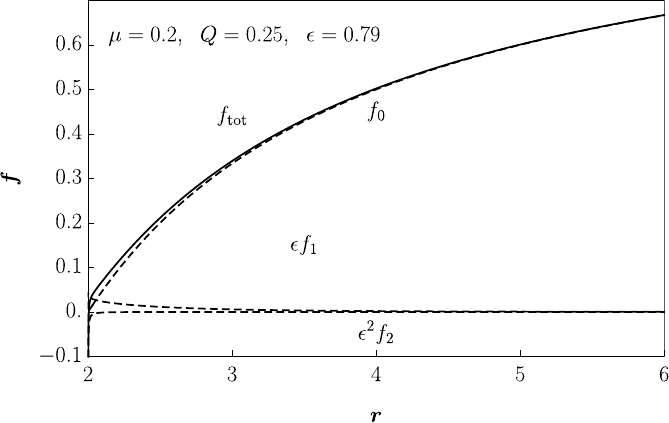}
\includegraphics[width=.49\linewidth]{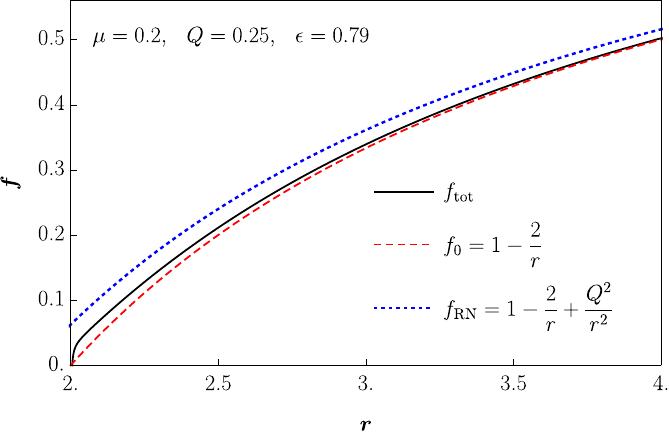}
\caption{The contribution of each individual terms of laps function to the total solution (left panel) and comparison with the solutions of Reissner Nordstrom, Schwardschild (right panel) for small $\epsilon=0.0315$ and $\epsilon=0.79$ in the case of constant $Q=0.25$ black hole charge.}
\label{fig_7}
\end{figure*}
\begin{figure*}[t]
\centering
\includegraphics[width=.49\linewidth]{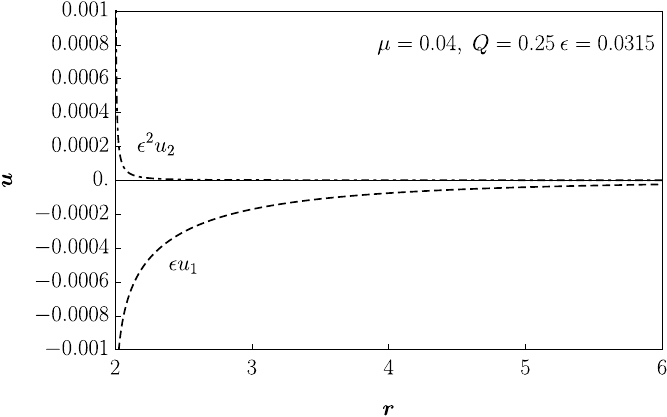}
\includegraphics[width=.49\linewidth]{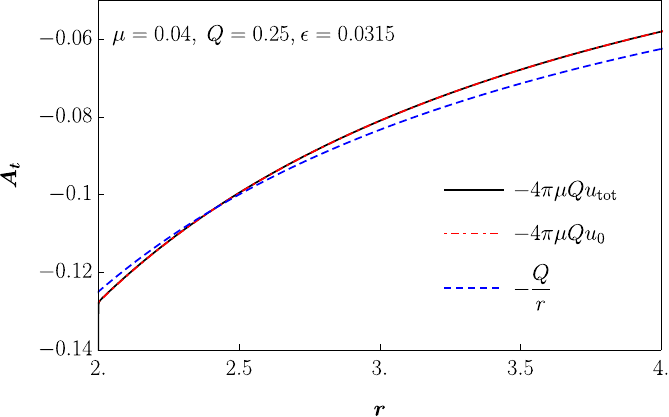}
\includegraphics[width=.49\linewidth]{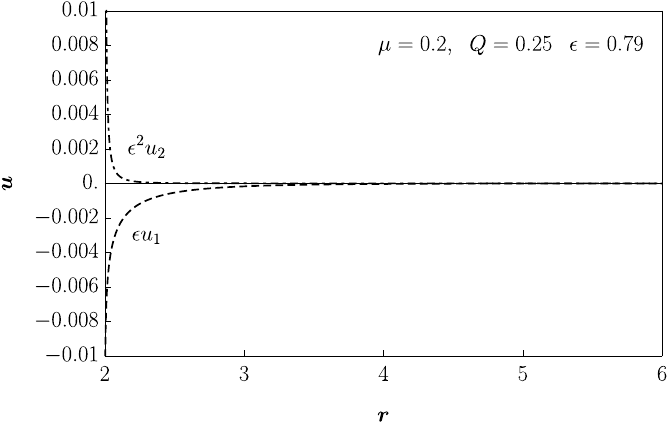}
\includegraphics[width=.49\linewidth]{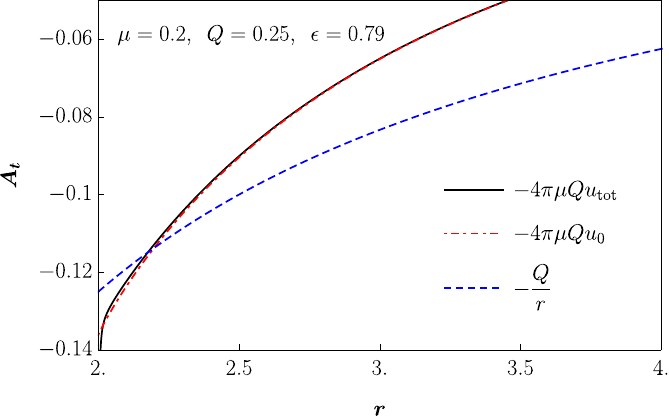}
\caption{The contribution of each individual terms of $u$ function to the total solution (left panel) and comparison with the Columb potential (right panel) for small $\epsilon=0.0315$ and $\epsilon=0.79$ in the case of constant $Q=0.25$ black hole charge. }
\label{fig_8}
\end{figure*}

\section{The contribution of high orders to the total solution}
Zero-order equations read 
\begin{eqnarray}
f_{0}\left(\frac{2f'_{0}}{x}+f''_{0}\right)&=&0,
\\
f_{0}\left(\frac{2u'_{0}}{x}+u''_{0}\right)&=&u_{0}\left(f_{0}+xf'_{0}\right),
\end{eqnarray}
which have exact solutions in an analytical form:
\begin{eqnarray}
    f_{0}&=&1-\frac{2C}{x},\\
    u_{0}&=&\frac{e^{-\left(x-2C\right)}}{4\pi x}U\left[C,0,2(x-2C)\right].
\end{eqnarray}
The asymptotic behaviour of a vector field at large distances can be described as follows 
\begin{equation}
     \lim\limits_{x \to \infty}u_{0}=\frac{e^{-(x-2C)}}{4\pi x}\left(\frac{1}{2x}\right)^{C}+\dots\equiv u_{0}^{\infty}+\dots
\end{equation}
After some mathematical simplifications, we can write an equation describing the second-order fields in the following form 
\begin{equation}
\frac{2f'_{1}}{x}+f''_{1}=\left(u_{0}'\right)^2+\frac{1}{2}\frac{u_{0}^2}{f_{0}^2}\big(4f_{0}+x f_{0}'\big). 
\end{equation}
Here, for simplicity, we omitted the equation for the vector field. It is important to note that the right-hand side of the equation diverges at $x=2C$ due to the factor $1/f_{0}^2$. Furthermore, the equation holds for $x>2C$ due to the presence of $u_{0}$, and therefore the first approximation should also exist between $2C<f_{1}<\infty$. The right-hand side of the equation is defined by analytic functions, which makes it possible to find boundary conditions. By analyzing the equation at infinity and near $x=2C$, we obtain  the following estimation 
\begin{eqnarray}
\label{f_1_near_hor}
     \lim\limits_{x \to 2C^{+}}f_{1}&=& -\frac{\ln{\left(x-2C\right)}}{2^{5}\pi^2 C^2\Gamma(C)^2} +\text{finite term}\\
    \lim\limits_{x \to \infty}f_{1}&=&3\left(\frac{u_{0}^{\infty}}{2}\right)^2+\dots 
\end{eqnarray}
Since in geometric units $C=M \mu$ together with $x=\mu r$, one can see that the function $f_{1}$ has a logarithmic divergence $f_{1}\rightarrow + \infty $ around $r=2M$ due to the divergent term in \eqref{f_1_near_hor}. In addition, one can see that these boundary conditions strongly depend on the parameter $C= \tilde{\mu}$, which plays the role of a massive parameter of the vector field. 

By analyzing the equations for the vector field, substituting the asymptotics of $f_{1}$ near $x=2C$ and at infinity, it is also possible to determine the boundary conditions for the first approximation of the vector field.
\begin{eqnarray}
\label{u_1_near_hor}
     \lim\limits_{x \to 2C^{+}}u_{1}&=& - \frac{\ln^2{\left(x-2C\right)}}{2^{6}\pi^3C^2\Gamma\left(C\right)^3}+\dots\\
     \lim\limits_{x \to \infty}u_{1}&=&-x\left(\frac{u_{0}^{\infty}}{2}\right)^3+\dots
\end{eqnarray}
These conditions demonstrate that the first-order vector field has a negative sign and also diverges logarithmically at the radius of $r=2M$.

By repeating the same process for the second-order approximation allows us to find the asymptotic behavior of functions $f_{2}$ and $u_{2}$. In conclusion, the asymptotic behavior of all terms at infinity in the vicinity of $r=2M$ can be written as 
\begin{strip}
\begin{eqnarray}
\label{F_1_infty} 
\lim\limits_{x \to \infty}f_{1}&=&3\left(\frac{u_{0}^{\infty}}{2}\right)^2+\dots\\
\label{U_1_infty} 
\lim\limits_{x \to \infty}u_{1}&=&-x\left(\frac{u_{0}^{\infty}}{2}\right)^3+\dots\\
\label{F_2_infty} 
\lim\limits_{x \to \infty}f_{2}&=&-6x\left(\frac{u_{0}^{\infty}}{2}\right)^{4}\left(1+2 C\right)+\dots\\ 
\label{U_2_infty} 
\lim\limits_{x \to \infty}u_{2}&=&x^2\left(\frac{u_{0}^{\infty}}{2}\right)^5\left(\frac{136+106C+405C^2+600C^3+405C^4}{96}\right)+\dots\\
 \lim\limits_{x \to 2C^{+}}2^{5}\pi^2 C^2\Gamma(C)^2f_{1}&=& -\ln{\left(x-2C\right)} +\text{finite terms}
 \\
 \lim\limits_{x \to 2C^{+}}2^{6}\pi^3C^2\Gamma\left(C\right)^3u_{1}&=&-\ln^2{\left(x-2C\right)}+\left(\frac{1-2C-4C^2W}{4C^2}\right)\ln\left(x-2C\right)+\text{finite terms}\\
    \lim\limits_{x \to 2C^{+}}2^{9}\pi^4C^3\Gamma(C)^4f_{2}&=&\frac{\ln{\left(x-2C\right)}}{x-2C}-2C\ln^3{\left(x-2C\right)}-\frac{5+8C-8C^2+16C^2W}{4C}\ln^2{\left(x-2C\right)}-\\
    &&\frac{1-4C+16C^3W+4C^2(5-2W)+16C^4(W^2-2)}{8C^3}\ln{\left(x-2C\right)}+\text{finite terms. }\nonumber
\end{eqnarray}
\end{strip}
Here we used 
\begin{equation} 
W=2\gamma+\ln{2}+\psi(C), 
\end{equation} 
where $\gamma$ and $\psi(C)$ are the Euler–Mascheroni constant and polygamma function, respectively. Using \eqref{F_1_infty}-\eqref{U_2_infty}, one can find a general expression at infinity for the $n$-th term in the form
\begin{eqnarray}
   \lim\limits_{x \to \infty}f_{n}&=&A_{n}(-1)^{n-1}x^{n-1}\left(\frac{u_{0}^{\infty}}{2}\right)^{2n}+\dots\\
   \lim\limits_{x \to \infty}u_{n}&=&B_{n}(-1)^{n}x^{n}\left(\frac{u_{0}^{\infty}}{2}\right)^{2n+1}+\dots,
\end{eqnarray}
where $A_{n}$ and $B_{n}$ are constants of the $n$-th term. 

In light of the preceding discussion, we can now present a summary of the key points.
\begin{itemize}
   \item 
   Higher-order terms contribute insignificantly to the complete solutions at large distances, where they are proportional to $f_{n}\approx (u_{\infty})^{2n}$, $u_{n}\approx (u_{\infty})^{2n+1}$ and thus can be neglected. Since circular orbits exist above the event horizon (at least $r>3M$ for small values of $\mu$), higher-order terms can be safely ignored, and only the zero-order approximation needs to be considered. 
  \item On the other hand, in the vicinity of the event horizon, $r=2M$, the relations $|f_{n+1} |>|f_{n}|$ and $|u_{n+1}|>|u_{n}|$ are satisfied. Therefore, the higher-order terms of $f_{n}$ and $u_{n}$ contribute more significantly to the complete solution near the radius $r=2M$. However, even in this case, the higher values are always compensated by the smallness of the parameter $\epsilon$. For further clarification, refer to the expansions of the functions in \eqref{u_f_expandsion}, see Figs. \ref{fig_7} and \ref{fig_8}.
\end{itemize}


\providecommand{\noopsort}[1]{}\providecommand{\singleletter}[1]{#1}%

\end{document}